\documentclass[prl,twocolumn,groupedaddress,showpacs,nofootinbib]{revtex4}
\usepackage{graphicx}
\usepackage{dcolumn}
\usepackage{bm}
\usepackage{amssymb}
\usepackage{mathrsfs}
\usepackage{amsmath}
\usepackage{epsfig}
\usepackage{here}
\usepackage[dvips]{color}
\usepackage{hhline}

\begin{document}

\title{Common Origin of Visible and Dark Universe}

\author{Pei-Hong Gu$^{1}_{}$}
\email{pgu@ictp.it}

\author{Utpal Sarkar$^{2}_{}$}
\email{utpal@prl.res.in}

\affiliation{$^{1}_{}$The Abdus Salam International Centre for
Theoretical Physics, Strada Costiera 11, 34014 Trieste, Italy\\
$^{2}_{}$Physical Research Laboratory, Ahmedabad 380009, India}

\begin{abstract}

Dark matter, baryonic matter and dark energy have different
properties but contribute comparable energy density to the present
Universe. We point out that they may have a common origin. As the
dark energy has a scale far lower than all known scales in particle
physics but very close to neutrino masses, while the excess matter
over antimatter in the baryonic sector is probably related to the
neutrino mass-generation, we unify the origin of dark and visible
Universe in a variant of seesaw model. In our model (i) the dark
matter relic density is a dark matter asymmetry emerged
simultaneously with the baryon asymmetry from leptogenesis; (ii) the
dark energy is due to a pseudo-Nambu-Goldstone-Boson associated with the
neutrino mass-generation.

\end{abstract}


\maketitle

Cosmological observations indicate that dark and visible matter
contribute comparable energy density to the present Universe
\cite{amsler2008}. This coincidence implies that the dark and
visible matter may have a common origin although their creation and
evolution are usually understood by unrelated mechanisms. The
visible matter exists in the present Universe as a
matter--antimatter asymmetry, which is the same as the baryon
asymmetry. If the amount of baryon and antibaryon were the same,
they would have annihilated and we could not have existed. The most
promising mechanism for generating a baryon asymmetry is
leptogenesis \cite{fy1986,lpy1986}, in which a lepton asymmetry is
first produced, which is then partially converted to a baryon
asymmetry by the sphaleron \cite{krs1985} process before the
electroweak phase transition.

In most models of dark matter, ones assume the dark matter to be a
neutral particle without any quantum number, and then adjust its
decay or annihilation rate to give a required relic density. There
is another possibility that the dark matter actually carry some
$U(1)$ quantum number so that the dark antimatter also exists. In
this scenario, the excess of dark matter over dark antimatter could
determine the amount of dark matter relic density if the dark matter
and dark antimatter have very fast annihilation rate. This mechanism
can explain the current existence of dark matter without fine tuning
of the decay or annihilation rate of dark matter. The dark matter
asymmetry could then be generated by the same mechanism that
generates the baryon asymmetry
\cite{kuzmin1997,kl2004,cht2005,klz2009,gsz2009}. Therefore, the
visible and dark matter have a common origin which automatically
implies their comparable energy density in the present Universe.

Cosmological observations also provide strong evidence that our
Universe is expanding with an accelerated rate. This acceleration
can be attributed to the existence of dark energy. It is striking
that the dark energy scale is far lower than all known scales in
particle physics except the neutrino masses. This coincidence may
have its origin in the neutrino dark energy model
\cite{gwz2003,bgwz2003}, where the dark energy is given by the
pseudo-Nambu-Goldstone-bosons (pNGB) associated with the neutrino
mass-generation \cite{bhos2005,ghs2007}.

In this paper, we propose a variant of seesaw
\cite{minkowski1977,mw1980} model to give a common origin for the
visible and dark matter by creating the baryon and dark matter
asymmetries simultaneously. At the same time this model explains how
the dark energy is related to the neutrino masses, and hence,
contributes comparable energy density to the present Universe with
the visible and dark matter.

\vspace{7mm}

We start with a simple version of our model to focus on the common
origin of the dark and baryonic matter. For simplicity, we do not
write down the full Lagrangian, which is supposed to be invariant
under a global symmetry of lepton number, instead we only give the
part that is relevant for our discussions,
\begin{eqnarray}
\label{yukawa} \mathcal{L_{\textrm{Y}}^{}} &\supset&
-\frac{1}{2}y\overline{\psi_{L}^{c}}
i\tau_2^{}\xi\psi_{L}^{}+\textrm{H.c.}\,,\\
[3mm]\label{potential} V&\supset&
M_\eta^2\left(\eta^\dagger_{}\eta\right)+m_\xi^2\textrm{Tr}\left(\xi^\dagger_{}\xi\right)+m_\chi^2\left(\chi^\dagger\chi\right)\nonumber\\
[2mm]
&&+\left(\kappa\eta\phi^{T}_{}i\tau_2^{}\xi\phi+\lambda\eta\chi^3_{}+\rho\eta\sigma^2_{}+\textrm{H.c}\right)\nonumber\\
[2mm]
&&+\left(\chi^\dagger_{}\chi\right)\left[\alpha\textrm{Tr}\left(\xi^\dagger_{}\xi\right)
+\beta\left(\phi^\dagger_{}\phi\right)+\gamma\left(\sigma^\dagger_{}\sigma\right)\right]\nonumber\\
[2mm]
&&+\left(\sigma^\dagger_{}\sigma\right)\left[\zeta\textrm{Tr}\left(\xi^\dagger_{}\xi\right)
+\epsilon\left(\phi^\dagger_{}\phi\right)+\vartheta\left(\sigma^\dagger_{}\sigma\right)\right]\,.
\end{eqnarray}
Here $\psi_{L}^{}(\mathbf{1},\mathbf{2},-1/2)$ and
$\phi(\mathbf{1},\mathbf{2},-1/2)$, respectively, are the lepton and
Higgs doublets, $\xi(\mathbf{1},\mathbf{3},2)$ is the Higgs triplet,
$\sigma(\mathbf{1},\mathbf{1},0)$ and
$\chi(\mathbf{1},\mathbf{1},0)$ are the light singlets,
$\eta(\mathbf{1},\mathbf{1},0)$ denotes the heavy singlet [under the
SM gauge group $SU(3)_{c}^{}\times SU(2)_L^{} \times U(1)_{Y}^{}$].
Conventionally, we assign a lepton number $L=+1$ for the lepton
doublets $\psi_{L}^{}$ while $L=0$ for the Higgs doublet $\phi$. In
order to exactly conserve the lepton number, we further assign
$L=-2$ for the Higgs triplet $\xi$, $L=+2$ for the heavy singlet
$\eta$, $L=-1$ for the light singlet $\sigma$ and $L=-2/3$ for the
light singlet $\chi$. Moreover, we impose a discrete $Z_{3}^{}$
symmetry under which the light singlet $\chi$ is the unique field
with non-trivial transformation [$\chi \rightarrow
\exp(i\frac{2\pi}{3})\chi$] so that it is stable without nonzero
vacuum expectation value (VEV).

For appropriate choice of parameters, the light singlet $\sigma$
develops a VEV of about TeV, which breaks the global lepton number
symmetry spontaneously. This VEV will, in turn, induce a small VEV
of the heavy singlet $\eta$ \cite{ghsz2009},
\begin{eqnarray}
\langle\eta\rangle \simeq
-\frac{\rho\langle\sigma\rangle^{2}_{}}{M_{\eta}^2}~~\textrm{for}~~M_{\eta}^{}\gtrsim
\rho \gg \langle\sigma\rangle\,,
\end{eqnarray}
and hence a highly suppressed trilinear lepton number violating
coupling of the Higgs triplet $\xi$ to the Higgs doublet $\phi$,
\begin{eqnarray}
\mu= \kappa\langle\eta\rangle
\simeq-\kappa\frac{\rho\langle\sigma\rangle^{2}_{}}{M_{\eta}^{2}}\,.
\end{eqnarray}
Subsequently when the Higgs doublet $\phi$ develops a VEV to break
the local electroweak symmetry, the Higgs triplet $\xi$ will also
pick up a tiny VEV \cite{ghsz2009},
\begin{eqnarray}
\langle\xi\rangle \simeq
-\frac{\mu\langle\phi\rangle^{2}_{}}{m_{\xi}^2}~~\textrm{for}~~\mu\ll\langle\phi\rangle\lesssim
m_{\xi}^{}\,.
\end{eqnarray}
Clearly the suppressed coupling $\mu$ can guarantee the tiny VEV
$\langle\xi\rangle$ although $m_{\xi}^{}=\mathcal{O}(\textrm{TeV})$
is mildly bigger than $\langle\phi\rangle\simeq 174\,\textrm{GeV}$.
Therefore the neutrinos eventually obtain their small Majorana
masses through the Yukawa couplings of the Higgs triplet to the
lepton doublets,
\begin{eqnarray}
m_{\nu}^{}= y\langle\xi\rangle\,.
\end{eqnarray}

\vspace{7mm}

In our model, the heavy singlet $\eta$ has three decay channels:
\begin{eqnarray}
\eta\,\rightarrow\, \xi^\ast_{}\,\phi^\ast_{}\,\phi^\ast_{} \,,\quad
\eta\,\rightarrow\, \chi^\ast_{}\chi^\ast_{}\chi^\ast_{}\,,\quad
\eta\,\rightarrow\, \sigma^\ast_{}\,\sigma^\ast_{} \,.
\end{eqnarray}
If the CP is not conserved, the above decays and their CP-conjugate
can generate a lepton asymmetry stored in the Higgs triplet $\xi$,
in the light singlet $\chi$ and in the light singlet $\sigma$,
respectively, after the heavy singlet $\eta$ goes out of
equilibrium. As the fields $\xi$, $\chi$ and $\sigma$ carry
nonvanishing lepton numbers, these fields would store different
types of lepton asymmetry. These three types of lepton asymmetry
would decouple from each other as they are produced, although the
total lepton asymmetry will be zero as a result of the exactly
lepton number conservation.
The generic features of these asymmetries are:\\
\begin{enumerate}
\item[1.] There is an asymmetry between the light singlet $\chi$
and its CP-conjugate. This asymmetry will survive since $\chi$ is
not related to other lepton number violating interactions. We will
show later this asymmetry can serve as the dark matter asymmetry to
give a desired dark matter relic density.
\item[2.] The lepton asymmetry in the Higgs triplet $\xi$ can be rapidly
transferred to a lepton asymmetry in the lepton doublets
$\psi_L^{}$, as the lepton number conserving decay of $\xi$ into two
$\psi_L^{}$ is in equilibrium at this time. After the lepton number
is spontaneously broken at the TeV scale, there will be a lepton
number violating interaction of the Higgs triplet coupling to the
Higgs doublet. This lepton number violation is extremely weak so
that the induced lepton number violating processes will not go into
equilibrium until the temperature is well below the electroweak
scale. Therefore the lepton asymmetry stored in the lepton doublets
could be partially converted to a baryon asymmetry by the sphaleron
action before it is washed out by the lepton number violating
processes \cite{ghsz2009}. Clearly this is a leptogenesis picture.
\item[3.]
The lepton asymmetry in the light singlet $\sigma$ will not affect the
baryon asymmetry of the Universe as $\sigma$ does not take part in
the sphaleron process. Eventually we will have a relic
density of a singlet Majoron, the massless Nambu-Goldstone boson
corresponding to the global lepton number violation. This Majoron is
harmless since its component from the Higgs triplet $\xi$ is highly
suppressed by $\langle\xi\rangle/\langle\sigma\rangle$.
\end{enumerate}

\begin{figure*}
\vspace{9cm} \epsfig{file=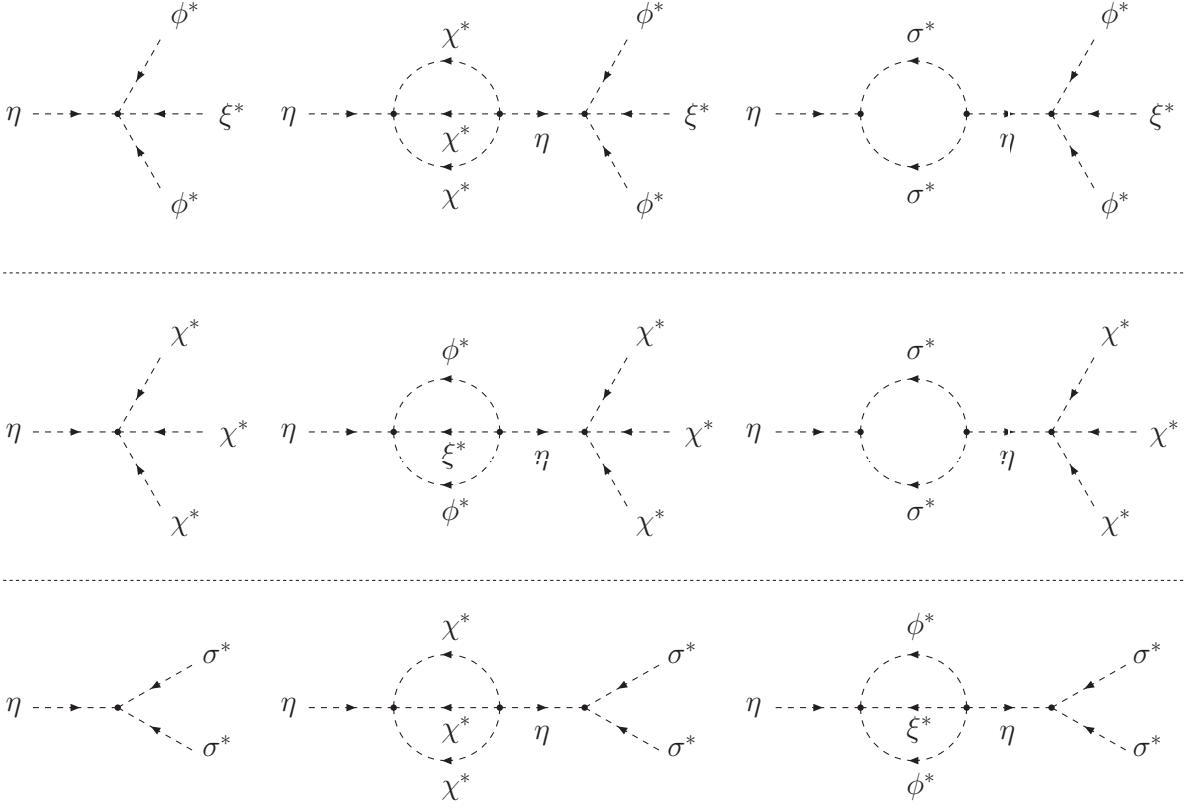, bbllx=6.3cm, bblly=5.9cm,
bburx=16.3cm, bbury=15.9cm, width=8cm, height=8cm, angle=0, clip=0}
\vspace{-5.4cm} \caption{\label{asymmetry} The lepton number
conserving decays of the heavy singlets $\eta$ at tree level and
loop orders for generating the lepton asymmetry stored,
respectively, in the Higgs triplet $\xi$, in the light singlet
$\chi$ and in the light singlet $\sigma$. The three types of lepton
asymmetry decouple from each other as they are produced. The lepton
asymmetry stored in the Higgs triplet $\xi$ will be partially
converted to the baryon asymmetry by the sphaleron process. On the
other hand, the asymmetry between the light singlet $\chi$ and its
CP-conjugate will survive as the relic density of dark matter.}
\end{figure*}

For realizing a CP violation in the decays of the heavy singlet
$\eta$, it is necessary that the tree-level diagrams interfere with
the self-energy loop diagrams as shown in Fig. \ref{asymmetry}. We
thus need at least two such heavy singlets $\eta$. Here we minimally
introduce two heavy singlets $\eta_{1,2}^{}$. For convenience, we
choose the base of $\eta_{1,2}^{}$ to give a real and diagonal mass
matrix $M_{\eta}^{2}=\textrm{diag}\left(M_{\eta_1^{}}^{2},
M_{\eta_2^{}}^{2}\right)$ and two real cubic couplings
$\rho=(\rho_{1}^{},\rho_{2}^{})$ by a proper rotation. Consequently
we only need to ensure $\kappa=(\kappa_1^{},\kappa_2^{})$ and
$\lambda=(\lambda_1^{},\lambda_2^{})$ to be complex. In the limiting
case where the two heavy singlets $\eta_{1,2}^{}$ have hierarchical
masses, the final lepton asymmetry stored in the Higgs triplet $\xi$
and the dark matter asymmetry stored in the light singlet $\chi$
should mainly come from the decays of the lighter one. For
illustration, let us focus on this hierarchical case. Without loss
of generality, we choose $\eta_{1}^{}$ to be the lighter heavy
singlet and $\eta_{2}^{}$ the heavier one. We then calculate the
lepton asymmetry stored in the Higgs triplet $\xi$ and the dark
matter asymmetry stored in the light singlet $\chi$ from the decays
of the lighter heavy singlet $\eta_{1}^{}$. For simplicity, we
assume
\begin{eqnarray}
\frac{\kappa_{i}^{}}{|\kappa_{i}^{}|}\equiv
\frac{\lambda_{i}^{}}{|\lambda_{i}^{}|}=e^{i\delta_i^{}}_{}\,,
\end{eqnarray}
and define $\delta\equiv \delta_{2}^{}-\delta_{1}^{}$, so that we
can easily read,
\begin{eqnarray}
\varepsilon_{\eta_1^{}}^{L_{SM}^{}}&\equiv&
2\frac{\Gamma(\eta_{1}^{} \rightarrow
\xi^\ast_{}\phi^\ast_{}\phi^\ast_{})-\Gamma(\eta_{1}^{\ast}\rightarrow\xi\phi\phi)}{\Gamma_{\eta_1^{}}^{}}\nonumber\\
[3mm]
&=&\frac{\sin\delta}{2\pi}\left|\frac{\kappa_{2}^{}}{\kappa_{1}^{}}\right|
\frac{\rho_{1}^{}\rho_{2}^{}}{M_{\eta_2^{}}^{2}-M_{\eta_1^{}}^{2}}\nonumber\\
[3mm] &&\times
\frac{\frac{3}{32\pi^{2}_{}}|\kappa_{1}^{}|^{2}_{}}{\frac{\rho_{1}^{2}}{M_{\eta_{1}^{}}^{2}}
+\frac{3}{32\pi^{2}_{}}\left(|\kappa_{1}^{}|^{2}_{}+|\lambda_{1}^{}|^{2}_{}\right)}\,,\\
[4mm] \varepsilon_{\eta_1^{}}^{\chi}&\equiv&
3\frac{\Gamma(\eta_{1}^{\ast}\rightarrow\chi\chi\chi)-\Gamma(\eta_{1}^{}
\rightarrow
\chi^\ast_{}\chi^\ast_{}\chi^\ast_{})}{\Gamma_{\eta_{1}^{}}^{}}\nonumber\\
[3mm]
&=&-\frac{3\sin\delta}{4\pi}\left|\frac{\lambda_{2}^{}}{\lambda_{1}^{}}\right|
\frac{\rho_{1}^{}\rho_{2}^{}}{M_{\eta_2^{}}^{2}-M_{\eta_1^{}}^{2}}\nonumber\\
[3mm] &&\times
\frac{\frac{3}{32\pi^{2}_{}}|\lambda_{1}^{}|^{2}_{}}{\frac{\rho_{1}^{2}}{M_{\eta_{1}^{}}^{2}}
+\frac{3}{32\pi^{2}_{}}\left(|\kappa_{1}^{}|^{2}_{}+|\lambda_{1}^{}|^{2}_{}\right)}\,.
\end{eqnarray}
It is straightforward to find the ratio between
$\varepsilon_{\eta_1^{}}^{L_{SM}^{}}$ and
$\varepsilon_{\eta_1^{}}^{\chi}$,
\begin{eqnarray}
\varepsilon_{\eta_1}^{L_{SM}^{}}:\varepsilon_{\eta_1}^{\chi}=|\kappa_{1}^{}\kappa_{2}^{}|:-\frac{3}{2}|\lambda_{1}^{}\lambda_{2}^{}|\,.
\end{eqnarray}
Here $\Gamma_{\eta_i^{}}^{}$ denotes the total decay width of
$\eta_{i}^{}$ or $\eta_{i}^{\ast}$,
\begin{eqnarray}
\Gamma_{\eta_i^{}}^{}&\equiv&\Gamma(\eta_{i}^{\,\,\,}\rightarrow
\xi^\ast_{}\phi^\ast_{}\phi^\ast_{})+\Gamma(\eta_{i}^{\,\,\,}\rightarrow
\chi^\ast_{}\chi^\ast_{}\chi^\ast_{})\nonumber\\
[2mm] &&+\Gamma(\eta_{i}^{}\, \rightarrow
\sigma^\ast_{}\sigma^\ast_{})\nonumber\\
[3mm]
&\equiv&\Gamma(\eta_{i}^{\ast}\rightarrow\xi^{\,\,\,}_{}\phi^{\,\,\,}_{}\phi^{\,\,\,}_{})
+\Gamma(\eta_{i}^{\ast}\rightarrow\chi^{\,\,\,}_{}\chi^{\,\,\,}_{}\chi^{\,\,\,}_{})\nonumber\\
[2mm] &&+\Gamma(\eta_{i}^{\ast}
\rightarrow \sigma^{\,\,\,}_{}\sigma^{\,\,\,}_{}) \nonumber\\
[3mm]
&=&\frac{1}{8\pi}\left[\frac{\rho_{i}^{2}}{M_{\eta_{i}^{}}^{2}}+\frac{3}{32\pi^{2}_{}}\left(|\kappa_{i}^{}|^{2}_{}
+|\lambda_{i}^{}|^{2}_{}\right)\right]M_{\eta_{i}^{}}^{}\,,
\end{eqnarray}
where the second equality is guaranteed by the unitarity and the CPT
conservation.

For generating a lepton asymmetry stored in the Higgs triplet $\xi$
and a dark matter asymmetry stored in the light singlet $\chi$, the
decaying particle $\eta_1^{}$ should match the out-of-equilibrium
condition \cite{kt1990}. For simplicity, we will consider the weak
washout regime with
\begin{eqnarray}
\Gamma_{\eta_1^{}}^{}\lesssim H(T)\left|_{T\simeq
M_{\eta_1}^{}}^{}\right.\,,
\end{eqnarray}
where
\begin{eqnarray}
H(T)&=&\left(\frac{8\pi^{3}_{}g_{\ast}^{}}{90}\right)^{\frac{1}{2}}_{}\frac{T^{2}_{}}{M_{\textrm{Pl}}^{}}
\end{eqnarray}
is the Hubble constant with the relativistic degrees of freedom
$g_{\ast}^{}\simeq 100$ and the Planck mass
$M_{\textrm{Pl}}^{}\simeq 10^{19}_{}\,\textrm{GeV}$. Furthermore,
there will emerge a lepton number violating interaction between the
Higgs triplet $\xi$ and the Higgs doublet $\phi$ when the light
singlet $\sigma$ develops a VEV $\langle\sigma\rangle$ to
spontaneously break the lepton number. The phase transition may
occur at the temperature $T_{c}^{}\lesssim \langle\sigma\rangle$. If
the induced cubic coupling $\mu$ between the Higgs scalars is highly
suppressed, specifically is much smaller than the mass of the Higgs
triplet $\xi$, i.e. $\mu\ll m_{\xi}^{}$, we have
\begin{eqnarray}
\Gamma(\xi\rightarrow
\phi^\ast_{}\phi^\ast_{})=\frac{1}{16\pi}\frac{|\mu|^{2}_{}}{m_{\xi}^{}}\ll
H(T)\left|^{}_{T=m_{\xi}^{}}\right.\,,
\end{eqnarray}
so the lepton number violating processes can only go into
equilibrium at a very low temperature and can not wash out the
lepton asymmetry stored in the Higgs triplet $\xi$ during the
sphaleron epoch. The Higgs triplet $\xi$ can thus transfer its
lepton asymmetry to the lepton doublets $\psi_{L}^{}$. Subsequently
the sphaleron process will partially convert this lepton asymmetry
to a baryon asymmetry.

\vspace{7mm}

The final baryon asymmetry and dark matter asymmetry would
contribute energy density to the present Universe as below
\cite{kt1990},
\begin{eqnarray}
\rho_{B}^{0}&=&n_{B}^{0}m_{N}^{}=\frac{n_{B}^{0}}{s_0^{}}m_{N}^{}s_0^{}
=-\frac{28}{79}\frac{n_{L_{SM}^{}}^{}}{s}\left|_{T\simeq
M_{\eta_1^{}}^{}}^{}m_{N}^{}s_0^{}\right. \nonumber\\
[2mm] &\simeq&
-\frac{28}{79}\varepsilon_{\eta_1^{}}^{L_{SM}^{}}\frac{n_{\eta_1^{}}^{eq}}{s}\left|_{T\simeq
M_{\eta_1^{}}^{}}^{}m_{N}^{}s_0^{}\right.\,,\\
[3mm]
\rho_{\chi}^{0}&=&n_{\chi}^{0}m_{\chi}^{}=\frac{n_{\chi}^{0}}{s_0^{}}m_{\chi}^{}s_0^{}
=\frac{n_{\chi}^{}}{s}\left|_{T\simeq
M_{\eta_1^{}}^{}}^{}m_{\chi}^{}s_0^{}\right. \nonumber\\
[2mm] &\simeq&
\varepsilon_{\eta_1^{}}^{\chi}\frac{n_{\eta_1^{}}^{eq}}{s}\left|_{T\simeq
M_{\eta_1^{}}^{}}^{}m_{\chi}^{}s_0^{}\right.\,.
\end{eqnarray}
Conventionally, we define
\begin{eqnarray}
\eta_{B}^{0}&=&\frac{n_B^{0}}{n_{\gamma}^{0}}\simeq7.04\times\frac{n_B^{0}}{s_{0}^{}}\nonumber\\
[2mm]
&\simeq&7.04\times\left[-\frac{28}{79}\varepsilon_{\eta_1^{}}^{L_{SM}^{}}\frac{n_{\eta_1^{}}^{eq}}{s}\left|_{T\simeq
M_{\eta_1^{}}^{}}^{}\right.\right]\nonumber\\
[2mm]
&\simeq&-\frac{7.04\,\varepsilon_{\eta_1^{}}^{L_{SM}^{}}}{15\,g_\ast^{}}\,,
\end{eqnarray}
to describe the current baryon asymmetry. Here $m_{N}^{}\simeq
1\,\textrm{GeV}$ is the masses of the nucleons, $s$ is the entropy
density, $n_{B}^{}$, $n_{\chi}^{}$ and $n_{\gamma}^{}$,
respectively, are the number density of baryon, dark matter and
photon, $n_{\eta_{1}^{}}^{eq}$ is the equilibrium distribution of
the heavy singlet $\eta_{1}^{}$. In the presence of fast
annihilation between the dark matter and dark anti-matter, the dark
matter asymmetry should be equivalent to the dark matter relic
density. In this scenario, the contributions from the baryonic and
dark matter to the present Universe should have the following ratio,
\begin{eqnarray}
\Omega_{B}^{}:\Omega_{\chi}^{}\equiv\rho_{B}^{0}:\rho_{\chi}^{0}=-\frac{28}{79}\varepsilon_{\eta_1^{}}^{L_{SM}^{}}
m_{N}^{}: \varepsilon_{\eta_1^{}}^{\chi} m_{\chi}^{}\,.
\end{eqnarray}

For giving a numerical example, we take
\begin{eqnarray}
&M_{\eta_1^{}}=0.1M_{\eta_2^{}}^{}=4\times
10^{13}_{}\,\textrm{GeV}\,,&\nonumber\\
[2mm]
&\rho_1^{}=\rho_2^{}=1.5\times 10^{12}_{}\,\textrm{GeV}\,,~~\langle\sigma\rangle=1\,\textrm{TeV}\,,&\nonumber\\
[2mm]
&m_{\xi}^{}=540\,\textrm{GeV}\,,~~m_{\chi}^{}=7\,\textrm{GeV}\,,&\nonumber\\
[2mm]
&|\kappa_{1}^{}|=|\kappa_{2}^{}|=2.4|\lambda_{1}^{}|=2.4|\lambda_{2}^{}|=1\,,&\nonumber\\
[2mm] &~~y= \mathcal{O}(1)\,,~~\sin\delta=-0.075\,,&
\end{eqnarray}
to output
\begin{eqnarray}
&\langle\chi\rangle \simeq -0.94\,\textrm{eV}\,,~~\mu\simeq
-0.94\,\textrm{eV}\,,~~\langle\xi\rangle\simeq 0.1\,\textrm{eV}\,,&\nonumber\\
[2mm] &\varepsilon_{\eta_1^{}}^{L_{SM}^{}}\simeq -4\,
\varepsilon_{\eta_1^{}}^{\chi}\simeq -1.4\times 10^{-7}_{}\,,&
\end{eqnarray}
and then find
\begin{eqnarray}
m_{\nu}^{}\sim 0.1\,\textrm{eV}\,,~~\eta_{B}^{0}\simeq 6.2\times
10^{-10}\,,~~\Omega_{\chi}^{}:\Omega_{B}^{}\simeq 5\,,
\end{eqnarray}
which are well consistent with the experimental observations
\cite{amsler2008}.

We now check if the annihilation between the dark matter and dark
anti-matter is so fast that the dark matter relic density can be
determined by the dark matter asymmetry. By taking into account that
$\langle\sigma\rangle=\mathcal{O}(\textrm{TeV})$ and
$\langle\phi\rangle\simeq 174\,\textrm{GeV}$, the thermally averaged
cross section in the non-relativistic limit is easy to read,
\begin{eqnarray}
\langle\sigma v\rangle&
=&\frac{1}{32\pi}\left[3\left(\alpha-\frac{\gamma\zeta}{2\vartheta}\right)^{2}_{}+2\left(\beta-\frac{\gamma\epsilon}{2\vartheta}\right)^2_{}
\right]\frac{1}{m_\chi^{2}}\nonumber\\
[2mm]
&&+\frac{1}{16\pi}\left(\frac{4}{9}-\frac{\gamma}{2\vartheta}\right)^2_{}\frac{m_\chi^2}{\langle\sigma\rangle^4_{}}\,\,
\textrm{for}\,\,m_\chi^{}=\mathcal{O}(\textrm{TeV})\,,\nonumber\\
\\
\langle\sigma v\rangle& =&
\frac{\beta^2_{}}{4\pi}\sum_{f}^{}N_f^c\frac{m_f^2}{m_h^4}\left(\frac{m_\chi^2-m_f^2}{m_\chi^2}\right)^{\frac{3}{2}}_{}
\,\,\textrm{for}\,\,m_\chi^{}=\mathcal{O}(\textrm{GeV})\,.\nonumber
\end{eqnarray}
Here $f$ denotes the SM fermions with $m_f^{}<m_\chi^{}$, $N_f^c$ is
the number of colors of the $f$-fermion, $h$ is the physical Higgs
boson defined by $\phi= \frac{1}{\sqrt{2}}h+\langle\phi\rangle$. By
inputting
$\alpha,\beta,\gamma,\zeta,\epsilon,\vartheta<\sqrt{4\pi}$, the
thermally averaged cross section is flexible to reach a large value.
For example, we obtain
\begin{eqnarray}
\langle\sigma v\rangle&
=&22\,\textrm{pb}\,\left(\frac{1\,\textrm{TeV}}{m_{\chi}^{}}\right)^2_{}\,,\nonumber\\
\langle\sigma v\rangle& =& 20\,\textrm{pb}\,
\left(\frac{7\,\textrm{GeV}}{m_{\chi}^{}}\right)^2_{}\left(\frac{120\,\textrm{GeV}}{m_h^{}}\right)^4_{}\,.
\end{eqnarray}
for $\alpha,\beta,\gamma=2$ and $\zeta,\epsilon,\vartheta=1$. It is
well known that the thermally produced dark matter with a mass from
a few GeV to a few TeV should have a thermally averaged cross
section slightly smaller than $1\,\textrm{pb}$ to give a right relic
density. If the thermally averaged cross section is too big, the
relic density will be much below the desired value. This means in
the present model, the thermally produced relic density is
negligible so that the dark matter asymmetry can naturally be a very
good approximation of the total relic density.

\vspace{7mm}

In the present scenario, our Universe will have mostly visible and
dark matter and very little visible or dark antimatter, which is the
main consequence of the models with common origin of visible and
dark matter through their asymmetries. This means that the
absence of decay or self-annihilation of dark matter will not
result in the overclosure of the Universe. In the absence of the
dark antimatter, the annihilation between the dark matter and dark
antimatter can not leave any significant products although the cross
section is very large. In this sense, the observed cosmic
positron/electron excess
\cite{chang2008,torii2008,adriani2008,aharonian2008,abdo2009} should
be from continuum distribution of pulsars
\cite{profumo2008,bgkms2009} or should have their origin in our
understanding of cosmic rays \cite{cowsik}. Consequently, the
gamma-ray radiation from the final states in the dark matter
annihilation is also absent. This is remarkably consistent with the
observations of the galactic center \cite{bergstrom2008} or the
center of dwarf galaxies \cite{essig2009}, which have already led to
strong constraints on the flux of gamma-ray radiation.

The dark matter scalar $\chi$ has a quartic coupling with the SM
Higgs doublet $\phi$, i.e.
$\beta\left(\chi^\dagger_{}\chi\right)\left(\phi^\dagger_{}\phi\right)$,
as given in Eq. (\ref{potential}). The induced cubic coupling is
\begin{eqnarray}
\label{dm-detect} V\supset \sqrt{2}\,\beta
\,\langle\phi\rangle\,h\,\left(\chi^\dagger_{}\chi\right) \,.
\end{eqnarray}
Through the s-channel exchange of the physical Higgs boson, the dark
matter is possible to find as a missing energy at colliders such as
the CERN LHC \cite{mcdonald1994}. On the other hand, the t-channel
exchange of the physical Higgs boson will result in an elastic
scattering of dark matter on nuclei and hence a nuclear recoil
\cite{mcdonald1994,aht2008}. The spin-independent cross section of
the dark-matter-nucleon elastic scattering would be,
\begin{eqnarray}
\label{dama3} \sigma\left(\chi N \rightarrow \chi N\right) & =&
\frac{\beta^2_{}}{4\pi}\frac{\mu_r^2}{m_h^4 m_\chi^2}f^2_{}m_N^2\,,
\end{eqnarray}
where $\mu_{r}^{}=m_\chi^{} m_N^{}/(m_\chi^{} + m_N^{})$ is the
nucleon-dark-matter reduced mass, $m_h^{}$ is the mass of the
physical Higgs boson, the factor $f$ in the range $0.14 <f<0.66$
with a central value $f=0.30$ \cite{aht2008} parameterizes the Higgs
to nucleons coupling from the trace anomaly,
$\displaystyle{fm_{N}^{}\equiv\langle
N|\sum_q^{}m_q^{}\bar{q}q|N\rangle}$. For the dark matter mass
within the range of 10 GeV to 1 TeV, we have
\begin{eqnarray}
&&\hskip -.5cm \sigma\left(\chi N \rightarrow \chi N\right)\simeq \nonumber\\
[3mm] && \hskip -.5cm \left[1.2\times
10^{-39}_{}\,\left(\frac{10\,\textrm{GeV}
}{m_\chi^{}}\right)^2_{}-1.7\times 10^{-43}_{}\,\left(\frac{
1\,\textrm{TeV}}{m_\chi^{}}\right)^2_{}\right]\textrm{cm}^2_{}\nonumber\\
[3mm] &&\hskip -.5cm \times
\frac{\beta^2_{}}{4\pi}\times\left(\frac{f}{0.3}\right)^2_{}
\times\left(\frac{120\,\textrm{GeV}}{m_h^{}}\right)^4_{}\,,
\phantom{xxx}
\end{eqnarray}
which could be naturally below the current experimental limit
\cite{angle2007,ahmed2008} and testable in the future experiments.
Recently, the DAMA collaboration \cite{bernabei2008} has observed an
annual modulation in the rates of nuclear recoil. If this signal is
confirmed, it should be induced by the scattering of the dark matter
particles from the galactic halo on the target nuclei in the
detectors. The good fitting \cite{pz2008} on the DAMA data and the
null results from other direct dark matter detection experiments
\cite{angle2007,ahmed2008} opens a small window for the
dark-matter-nuclei elastic scattering with the spin-independent
cross section and the dark matter mass as below,
\begin{eqnarray}
\label{dama1} &3\times 10^{-41}_{}\,\textrm{cm}^2_{}\lesssim
\sigma\lesssim 5\times 10^{-39}_{}\,\textrm{cm}^2_{}\,,&\\
[3mm] \label{dama2} &3\,\textrm{GeV}\lesssim m\lesssim
8\,\textrm{GeV}\,.&
\end{eqnarray}
In our model, the cross section (\ref{dama1}) can be easily matched
by inputting $\beta=\mathcal{O}(0.1-1)$, $0.14 <f<0.66$ and
$m_h^{}=120\,\textrm{GeV}$ with the mass (\ref{dama2}) to Eq.
(\ref{dama3}).

\vspace{7mm}

We now present the full version of our model to include the
coincidence between the dark energy and the neutrinos. For
simplicity, we only give the couplings relevant for our discussions,
\begin{eqnarray}
\label{lagrangian} \mathcal{L} &\supset&
-\sum_{i,j,k,\ell=1}^{3}\left[\frac{1}{2}y_{ij}^{}\overline{\psi_{L_i^{}}^{c}}
i\tau_2^{}\xi_{ij}^{}\psi_{L_j^{}}^{}+\kappa_{ij}^{}\eta_{ij}^{}\phi^{T}_{}i\tau_2^{}\xi_{ij}^{}\phi\right.\nonumber\\
[3mm]
&&+\lambda_{ij}^{}\eta_{ij}^{}\chi^3_{ij}+\omega_{ij}^{}\zeta_{ij}^{}\eta_{ij}^{}\sigma^2_{}
+\vartheta_{ijk\ell}\left(\zeta_{ij}^{\dagger}\zeta_{k\ell}^{}\right)\left(\eta_{ij}^{\dagger}\eta_{k\ell}^{}\right)\nonumber\\
[3mm] &&\left.+\textrm{H.c}\right]\,.
\end{eqnarray}
Here $\psi_{L}^{}$, $\phi$, $\xi$, $\eta$, $\sigma$ and $\chi$ keep
the definitions in the simple version. The six SM singlets
$\zeta_{ij}^{}=\zeta_{ji}^{}$ (without lepton number) have
independent phase transformations to give a global $U(1)^{6}_{}$
symmetry. In the presence of the Yukawa couplings of the Higgs
triplets to the lepton doublets (the first term in Eq.
(\ref{lagrangian})), this $U(1)^{6}_{}$ is explicitly broken down to
its subgroup $U(1)^{3}_{}$. So, there will emerge three massive
pNGBs associated with the neutrino mass-generation after the six
$\zeta_{ij}^{}$ acquire their VEVs. The Coleman-Weinberg effective
potential of these pNGBs would be
\begin{eqnarray}
V=-\frac{1}{32\pi^{2}_{}}\sum_{k=1}^{3}m_{k}^{4}\ln
\frac{m_{k}^{2}}{\Lambda^{2}_{}}\,,
\end{eqnarray}
where $m_{k}^{}$ as a function of the pNGBs is the \textit{k}th
eigenvalue of the neutrino mass matrix $m_{\nu}^{}$, and $\Lambda$
is the ultraviolet cutoff. A typical term in the potential $V$ has
the form,
\begin{eqnarray}
V(Q)\simeq V_{0}^{}\cos \left(\frac{Q}{M}\right)
\end{eqnarray}
where $Q$ is a pNGB combination with $M=\langle\zeta_{ij}^{}\rangle$
and $V_{0}^{}=\mathcal{O}(m_{\nu}^{4})$. It is well known that with
$M$ at the Planck scale $M_{\textrm{Pl}}^{}$, the pNGB field $Q$
will acquire a mass of the order of
$\mathcal{O}\left(m_{\nu}^{2}/M_{\textrm{Pl}}^{}\right)$ and thus
can provide a consistent candidate for the quintessence
\cite{weiss1987,wetterich1988} dark energy.

\vspace{7mm}

In this paper we propose a variant of seesaw model that provides a
common origin of the visible and dark matter and relates the dark
energy to the neutrino masses. The same origin automatically implies
that their contribution to the energy density of the present
Universe is comparable. In this model, there is a dark matter
asymmetry produced together with a lepton asymmetry that explains
the baryon asymmetry via the sphaleron process. This dark matter
asymmetry can account for the dark matter relic density because the
annihilation between the dark matter and dark antimatter is so fast
that the thermally produced relic density should be negligible.
Although the annihilation between the dark matter and dark
antimatter has a very large cross section, it is not required to
prevent the overclosure of the Universe. In the absence of dark
antimatter, currently the dark matter annihilation can't leave
significant products, which should then provide a theoretical
support to the alternate theories of the observed cosmic
positron/electron excess. The dark matter scalar has a quartic
coupling with the SM Higgs doublet so that it is expected to produce
at the colliders and/or detect by the direct dark matter detection
experiments. For example, the induced dark-matter-nucleon elastic
scattering can explain the DAMA signal and the null results from
other direct dark matter detection experiments. In our model, the
neutrino masses are functions of the dark energy field, which will
evolute with time and/or in space. In consequence, the neutrino
masses are variable, rather than constant. The prediction of the
neutrino-mass variation \cite{gwz2003} could be verified in the
experiments, such as the short gamma ray burst \cite{ldz2004}, the
cosmic microwave background and the large scale structures
\cite{bbmt2005}, the extremely high-energy cosmic neutrinos
\cite{rs2006} and the neutrino oscillations \cite{knw2004}.

\vspace{7mm}

\noindent \textbf{Acknowledgement}: PHG thanks Xinmin Zhang for
helpful discussions. US thanks the Department of Physics and the
McDonnell Center for the Space Sciences at Washington University in
St. Louis for inviting him as Clark Way Harrison visiting professor,
where part of this work was done.

\end{document}